\documentclass[conference]{IEEEtran}
\usepackage{geometry}
\usepackage{graphicx}
\begin{document}

\title{\Large\bf Defects Mitigation in Resistive Crossbars for Analog Vector Matrix Multiplication}

	\author{\normalsize
	\begin{tabular}[t]{c@{\extracolsep{8em}}c}
		\large Fan Zhang& \large Miao Hu \\
		\\
		Electrical and Computer Engineering & Electrical and Computer Engineering \\
		Binghamton University  & Binghamton Universiy \\
		Binghamton, New York~~13902 & Binghamton, New York~~13902\\
		e-mail: fzhang27@binghamton.edu & e-mail miaohu@binghamton.edu\\
\end{tabular}}


\maketitle

\makeatletter
\def\ps@IEEEtitlepagestyle{%
  \def\@oddfoot{\mycopyrightnotice}%
  \def\@evenfoot{}%
}
\makeatother
\def\mycopyrightnotice{%
  \begin{minipage}{\textwidth}
    \footnotesize
    978-1-7281-4123-7/20/\$31.00 ©2020 IEEE \hfill\\~\\
  \end{minipage}
  \gdef\mycopyrightnotice{}
}

{\small\bf Abstract---
With storage and computation happening at the same place, computing in resistive crossbars minimizes data movement and avoids the memory bottleneck issue.
It leads to ultra-high energy efficiency for data-intensive applications.
However, defects in crossbars severely affect computing accuracy. 
Existing solutions, including re-training with defects and redundant designs, but they have limitations in practical implementations. 
In this work, we introduce row shuffling and output compensation to mitigate defects without re-training or redundant resistive crossbars.
We also analyzed the coupling effects of defects and circuit parasitics.  
Moreover, We study different combinations of methods to achieve the best trade-off between cost and performance.
Our proposed methods could rescue up to 10\% defects in ResNet-20 application without performance degradation.}

\section{Introduction}

Memory bottleneck holds back data intensive applications, such as machine learning, image processing, and Internet-of-Things (IoT) \cite{wulf1995hitting}. 
This happens because of the limited on-chip memory resource and high cost off-memory access. 
As on-chip memory is way smaller than the data to be processed, frequent cache update with off-chip memory is necessary and it incurs significant energy consumption on off-chip communications, preventing today's computers from becoming more energy efficient.  
Recently, resistive crossbar-based computing attracts researchers' attention for its "in-memory computing" feature\cite{prezioso2015training,li2018analogue}. 
By storing matrix values non-volatilely as conductance of cross-point devices, a crossbar can efficiently perform vector-matrix multiplication(VMM) in one operation cycle \cite{yang2013memristive,Qiao:2018:AUR:3195970.3195998,Shafiee:2016:ICN:3007787.3001139,7551380}.
Using resistive crossbars, matrices only need to be fetched once, and the total communication cost dramatically reduces.  

However, defects in resistive crossbars is a major concern since they severely affect the VMM computing accuracy.
To mitigate defects, existing works use redundant circuits or re-training with consideration of defects \cite{liu2018design,kim2018energy,liu2017rescuing,chakraborty2017technology}, but they have certain limitations in practical applications.  
In this paper, we introduce row shuffling and output compensation to mitigate defects in resistive crossbars without need of redundant circuits or re-training. Our contributions are summarized as below:

\begin{itemize}
    \item Our defect mitigation methods do not need re-training and can apply to general VMM operations. With high flexibility, they could be used individually or in any combination. 
    \item We also discussed the coupling effect between circuit parasitic and defects, and we adopted parasitic-aware mapping to evaluate the performance of defect mitigation methods when circuit parasitic could not be ignored. 
    \item With experiment-verified circuit simulations, our proposed methods could rescue up to 10\% defects in crossbars on general VMM and modern CNN applications.
\end{itemize}

\section{Preliminary}
\subsection{Computing with resistive crossbar}
\begin{figure}[t]
	\centering
	\includegraphics[width=0.3\textwidth]{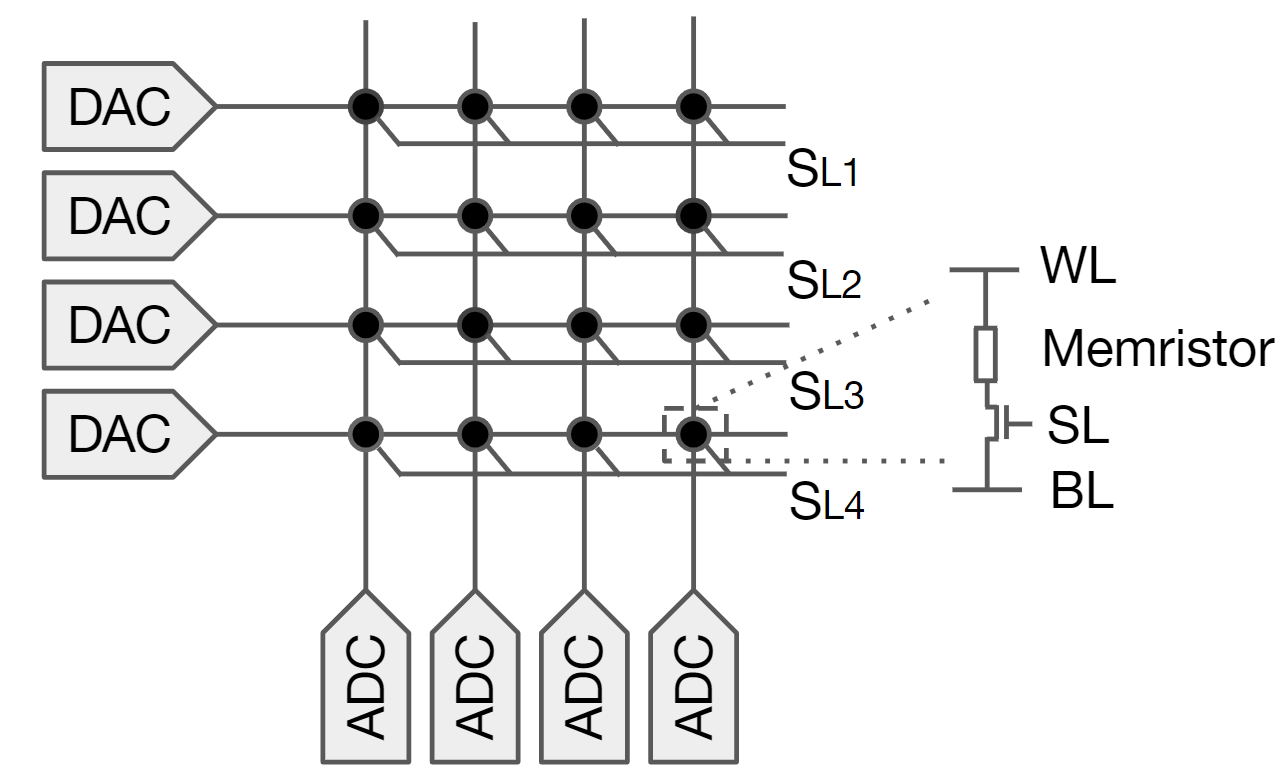}
	\vspace{-6pt}
	\caption{1-transistor-1-memristor (1T1M) crossbar for analog VMM computing with digital interface. \vspace{-6pt}}
	\vspace{-10pt}
	\label{fig_xbar_illus}
\end{figure}

In VLSI, the resistive crossbar is defined as an adjustable resistor array sandwiched by horizontal and vertical metal paths. 
The adjustable resistor could be the phase change material(PCM), the memristor, the floating gate,  the ReRAM device, the SRAM device, and etc. 
Fig. \ref{fig_xbar_illus} shows a diagram of the 1-transistor-1-memristor(1T1M) resistive crossbar array. 
It consists of wordlines(WL), bitlines(BL), selection lines(SL), access transistors and memristors.
The access transistor controlled by SL enables high precision, non-disturbance tuning for memristor conductance states\cite{merced2016repeatable}.
DAC and ADC arrays act as digital/analog interfaces for the crossbar to other digital components.
Researchers found that by mapping the matrix \textbf{A} to the conductance of cross-point device \textbf{G} and feed row vector \textit{X} as the input voltage signal \textit{V}, resistive crossbars can perform VMM via Ohm's Law and Kirchhoff's Current Law (KCL)\cite{hu2018memristor,hu2016dot}. 

\subsection{Impact of defects in resistive crossbars}
Unfortunately, defects exist in the resistive crossbar. Here, a defect is defined as a memristor whose conductance could not be programmed and stuck at certain state(s).  
There are various types of defects in a memristor crossbar\cite{ravi2017memristor}. 
Without losing generality, here we simplify divide defects into two most popular types: Stuck-ON defect at the highest conductance state, and Stuck-OFF defect at the lowest conductance state. 
Using a sequence of reads and writes, the traditional March test is enough to detect stuck-at faults\cite{6725492}.
    
\begin{figure}[t]
      \centering
      \includegraphics[width=0.8\columnwidth]{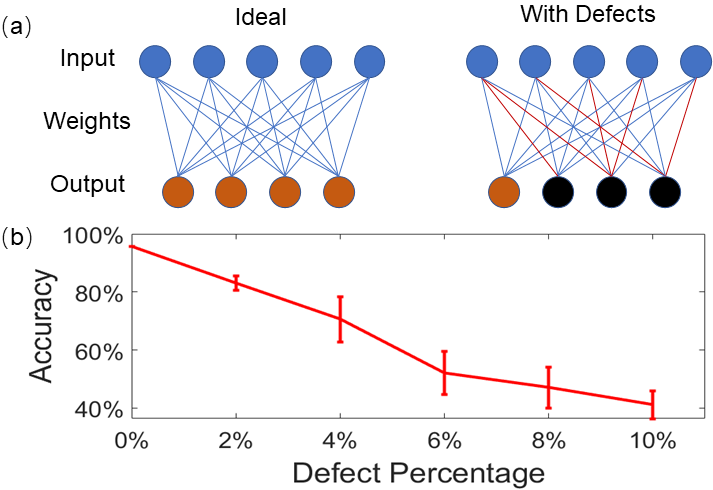}
      \vspace{-8pt}
      \caption{(a) An illustration of how defects affect ANN output. (b) ANN classification accuracy degrades as defect grows. \vspace{-6pt}}
      \label{fig_defects_illus}
\end{figure}

Defects significantly reduce computing accuracy of crossbars as they cause error to the target conductance values, and contribute to computing error in the output result.
In Fig. \ref{fig_defects_illus}(a), red lines between input and output neurons are representing weights stored on defected synapses. The output neurons (in black) receive contaminated weighted input signals from defected synapses, and it finally leads to a degradation in ANN performance.
classification accuracy drops as the defect rate grows \cite{lecun-mnisthandwrittendigit-2010}.
In Fig. \ref{fig_defects_illus}(b), the classification accuracy of ANN can reach around 95\% on MNIST without defects. 
After injecting 10\% defects\cite{6725492} in the weight matrices, classification accuracy quickly drops to around 50\%.
In short, crossbar-based ANN is sensitive to defects and it is necessary to mitigate defects in resistive crossbars.   

\section{Methodology}
\subsection{Row shuffling (RS)}
\begin{figure}[t]
      \centering
      \includegraphics[width=0.9\columnwidth]{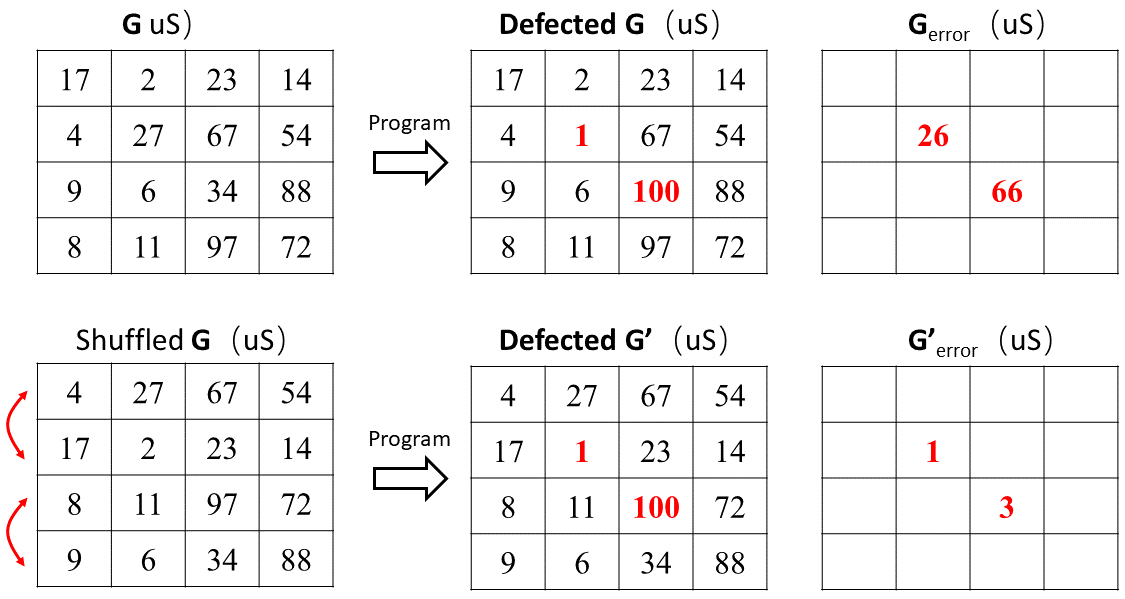}
      \vspace{-6pt}
      \caption{Shuffling rows in a $4\times4$ crossbar to minimize conductance error due to defects. \vspace{-6pt}}
      \label{fig_rowshuffle}
\end{figure} 
We observe that a defect becomes effective only when its stuck conductance being different from the target conductance. 
In other words, if we can arrange target conductance close to the stuck conductance of defect, the impact of the defect is reduced.
Fig. \ref{fig_rowshuffle} illustrates an example of RS to minimize conductance mapping error. 
By shuffling row 1 and row 2, row 3 and row 4 in \textbf{G}, new conductance mapping errors become smaller. 
Note that input channels to each row should also be shuffled accordingly. 
Although it is straightforward to shuffle rows in this simple example, it is not trivial to optimize RS in large crossbars with many defects.

Finding the best RS order in a defected crossbar can be defined as an assignment problem:
Before mapping \textbf{G} to a $n \times n$ crossbar having $m$ defected rows with index [$d_{1},d_{2},...,d_{m}$], the cost matrix \textbf{C} is generated. 
Each entry $c_{i,j}$ stores the total conductance error on each row due to defects (use L1 norm since errors could be positive or negative), when mapping the \textit{i}th row of \textbf{G} to the \textit{j}th row of a crossbar.
If there are no defects on a row, the cost of mapping any row of \textbf{G} to it is always 0. 
Our goal is to find an optimal set [$rs_1,rs_2,...,rs_m$], where the $rs_i$ row in \textbf{G} is shuffled to the $d_i$ row in crossbar, so that $\sum_{i=1}^{m} C_{rs_i,d_i}$ is minimized.   
The remaining $m-n$ rows in \textbf{G} can be mapped to the $m-n$ rows in crossbar.

To solve this problem, first, we scan (read-write-read) the whole crossbar to get its defect pattern. 
The conductance and locations of defects are both stored for later calculation.
Second, 
we convert the target matrix to the conductance matrix \textbf{G} 
so that we can generate the initial cost matrix \textbf{C}, where $c_{i,j}$ stores the L1 norm of the conductance error between the $i$th row of the crossbar and the $j$th row of \textbf{G}.
The conductance error is the difference between defect's conductance to the target conductance. 
In the end, We adopt Munkres algorithm to solve the assignment problem \cite{bourgeois1971extension}.

One limitation of RS is that the conductance mapping error is only minimized, but not completely removed.
However, RS always improves computing accuracy since it guarantees to minimize the overall conductance error due to defects, and it can be easily implemented by re-ordering input data.     

\subsection{Output compensation (OC)}
To further reduce the impact of defects, we estimate the error current caused by the defect and try to compensate it with affordable post-processing computation.
The post-processing computation must be low complexity and low cost to implement, otherwise, it is meaningless to fix defected crossbars with methods even more expensive than just re-calculating all with digital circuits.  

With this idea in mind, we estimate the error current due to defects in a simple yet effective way. 
In an ideal crossbar where ${\textit{I} = \textit{V}\textbf{G}}$, the error current caused by defects on the \textit{j}th column can be calculated by eq. \ref{ierror}.
\begin{equation}
I_{j,ideal}^{error} = \sum_{i = 1}^{k} (G_{l_i,j} - G_{l_i,j}^{defect}) \cdot v_{l_i}
\label{ierror}
\end{equation}
$v_{i}$ is the input voltage signal at the \textit{i}th row. 
\textit{k} ($k \ll n$) defects locate on the \textit{j}th column with index [$l_1,l_2,l_3,...,l_k$].
$G_{l_i,j}^{defect}$ is the conductance of defect on the $l_i$th row and the \textit{j}th column.

The actual error current on the \textit{j} column is formulated as below: 
\begin{equation}
I_{j,act}^{error} = \sum_{i=1}^{n} v_i \cdot G_{i,j} - I_{j,act} 
\label{ierror_actual}
\end{equation}
where $I_{j,act}$ is the actual output current sensed on the \textit{j}th column. 

\begin{figure}
    \centering
    \includegraphics[width=\columnwidth]{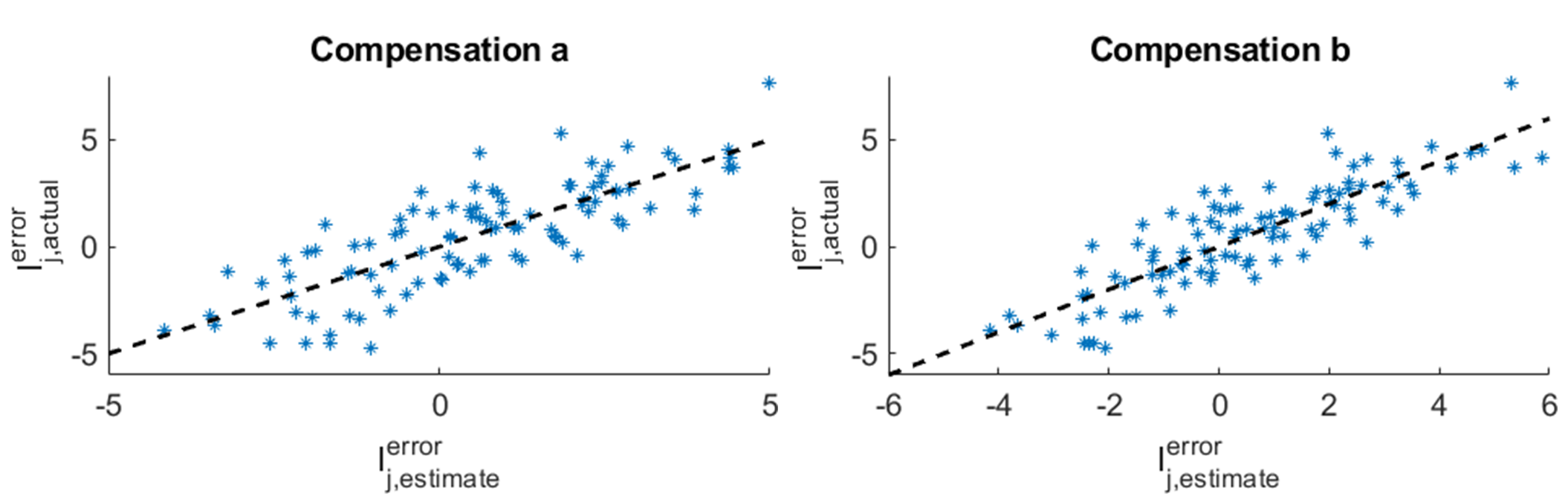}
    \vspace{-20pt}
    \caption{Fitting result with linear conductance mapping\vspace{-12pt}}
    \label{fig_m2ab}
\end{figure}

As running circuit simulation in real-time to fix crossbar output violates the rule of low complexity, we attempt to use the simple linear fitting method to estimate the actual error current with ideal error current. 
Linear fitting methods can be regarded as finding $a_j$ and $b_j$ coefficients so that $I_{j,est}^{error} \approx I_{j,act}^{error}$, where
\begin{equation}
I_{j,est}^{error} = a_j \cdot I_{j,ideal}^{error} + b_j  = \sum_{i = 1}^{k} a_j \cdot (G_{l_i,j} - G_{l_i,j}^{defect}) \cdot v_{l_i} + b_j
\label{ifit_a}
\end{equation}
Inspired by eq. \ref{ifit_a}, an alternative linear fitting method is given below to give more degree of freedom in fitting parameters, where 
\begin{equation}
{I'}_{j,est}^{error} = \sum_{i = 1}^{k} w_{l_i,j} \cdot v_{l_i} + b_j
\label{ifit_b}
\end{equation}
To distinguish the two fitting methods, we denote fitting with eq. \ref{ifit_a} as compensation a, and fitting with eq. \ref{ifit_b} as compensation b. 
Fig. \ref{fig_m2ab} shows the fitting result with both methods for a $128 \times 128$ crossbar. 
The simulation considers device models and wire resistances, and its setup is detailed in section \ref{sec_simulation}. Although a linear trend can be observed between $I_{j,est}^{error}$ and $ I_{j,act}^{error}$, the parasitic effect becomes unignorable, which causes large variations and reduces the quality and robustness of the linear fitting between the estimated error current to the real error current.
In short, it is necessary to compensate the impact of circuit parasitic to use OC in large-scale crossbars. 

\subsection{Parasitic-aware matrix mapping (PM)}

\begin{figure}[t]
      \centering
      \includegraphics[width=0.3\textwidth]{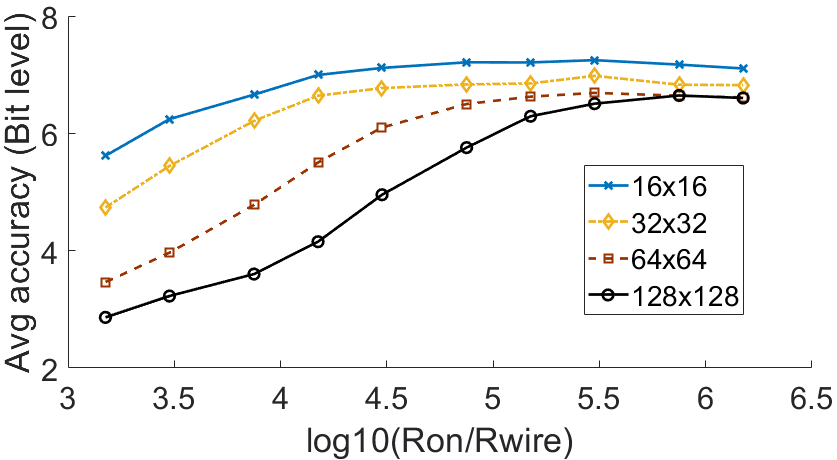}
      \vspace{-10pt}
      \caption{Impact of $R_{wire}$ on computing accuracy.  \vspace{-6pt}}
       \vspace{-10pt}
      \label{fig_Rw}
\end{figure}  

\begin{figure}
    \centering
    \includegraphics[width=\columnwidth]{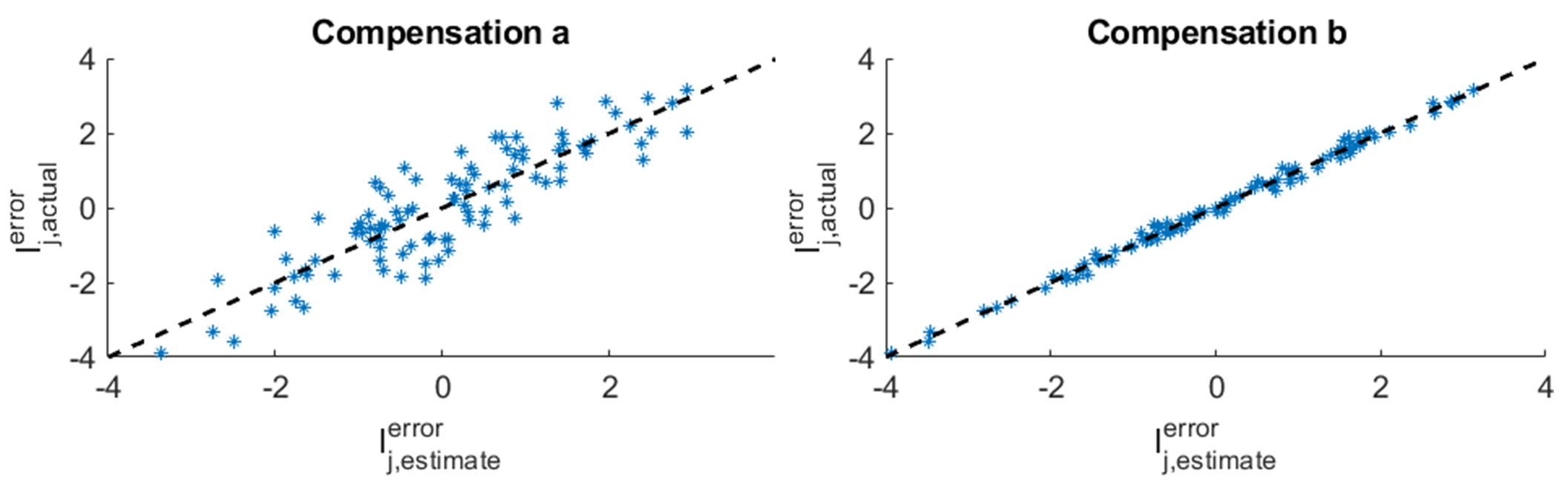}
    \vspace{-20pt}
    \caption{Fitting result with parasitic-aware mapping. \vspace{-6pt}}
    \label{fig_m2abm3}
\end{figure}

So far we are only using linear mapping to map \textbf{A} to \textbf{G} without considering the impact of circuit parasitics. However, in realistic, the impact of circuit parasitics, such as wire resistance and device nonlinearity, is usually ubiquitous and inevitable. 
Fig.\ref{fig_Rw} shows the impact of wire resistance on crossbar computing accuracy.
The circuit simulation is done with experiment-verified memristor model and transistor model\cite{hu2018memristor}. 
Linear mapping is used to map matrix value \textbf{A} to crossbar conductance \textbf{G}.  
$R_{on}$ is the lowest resistance state of the memristor device, and $R_{wire}$ is the interconnect resistance between adjacent cross-point devices.
As expected, computing accuracy improves with more conductive wires. 
However, the error does not decrease to 0 even with superconductive wires. This is because device nonlinearity starts to dominate the computing error when wire resistance can be ignored.   
One possible solution to eliminate such impact is to map \textbf{A} to \textbf{G} with consideration of wire resistance as well as device nonlinearity. Since mapping \textbf{A} to \textbf{G} only needs to be done once, it can afford additional computing cost to compensate parasitics in the mapping stage from \textbf{A} to \textbf{G}. 

We adopt the parasitic-aware mapping method from \cite{hu2016dot}, it targets on finding a new conductance matrix \textbf{G$'$} with consideration of parasitic effects, so that cross-point currents in a real crossbar with \textbf{G$'$} equal to the cross-point currents in an ideal crossbar with \textbf{G}.
To find \textbf{G$'$} for a $n \times n$ crossbar, additional constraints are added in circuit simulations to change devices to conductance states that can pass the ideal amount of current. 
For the detailed implementation of parasitic-aware mapping, please refer to \cite{hu2016dot}.

Fig. \ref{fig_m2abm3} shows the fitting result with parasitic-aware mapping instead of linear mapping. 
Comparing to Fig. \ref{fig_m2ab}, the quality of fitting significantly improves, especially for compensation b. This verifies that parasitic-aware mapping successfully decouples the parasitic and defects by mitigating parasitic effects. 
It also proves that the higher degree of freedom in compensation b helps and it should be our final version of output compensation. 

\begin{figure}[t]
      \centering
      \includegraphics[width=\columnwidth]{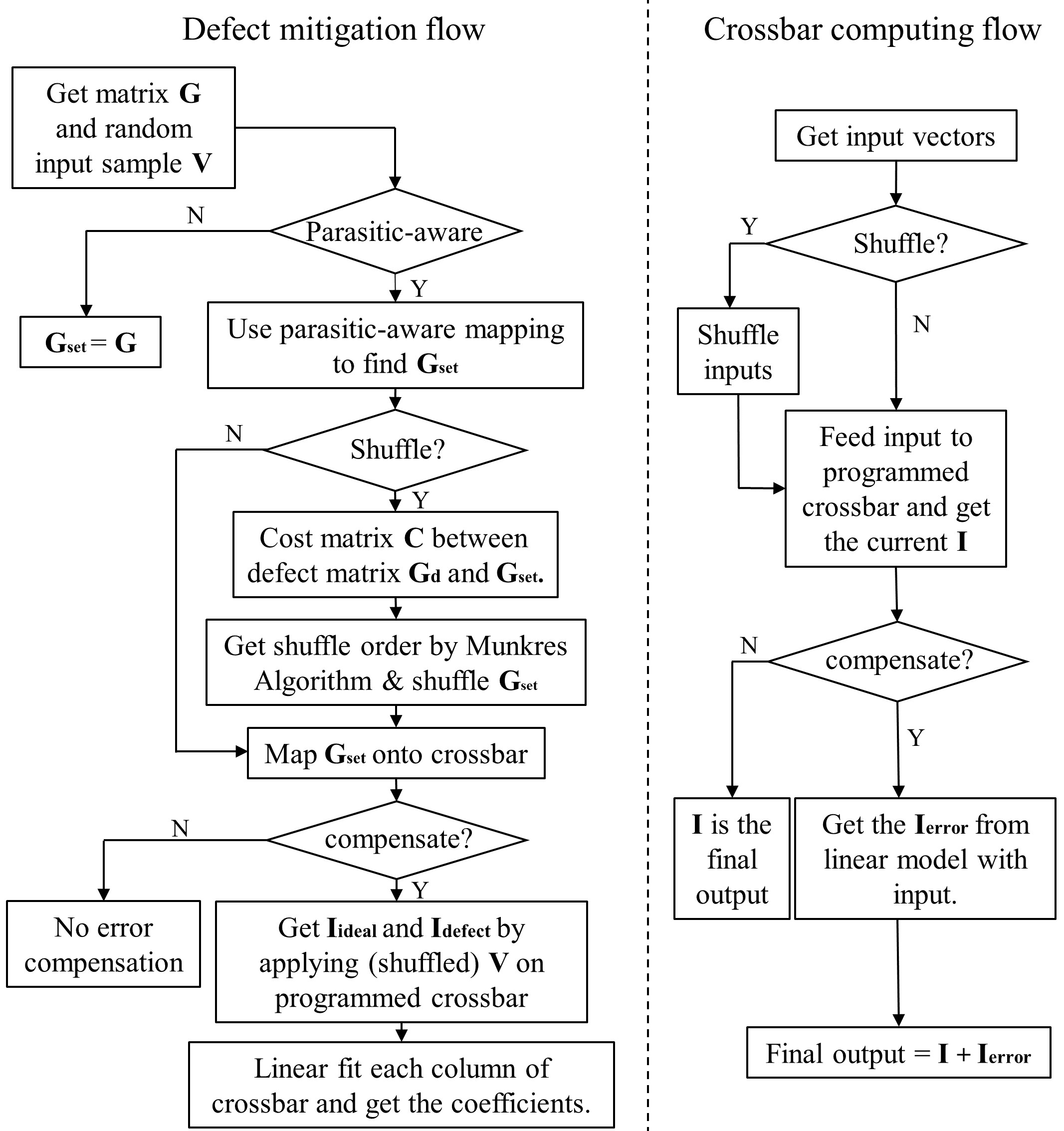}
      \vspace{-9pt}
      \caption{Defect mitigation flow chart and crossbar computing flow chart. \vspace{-8pt}}
      \vspace{-10pt}
      \label{fig_m1m2m3flow}
\end{figure}  

\subsection{General flow of the combined solution}
\label{subsec_summary}

Fig. \ref{fig_m1m2m3flow} summarizes the general flow of using row shuffling, output compensation, and parasitic-aware mapping in different combinations. 
Each method can be used independently and works on different stage. 
Parasitic-aware mapping takes more computations than linear mapping but offers better computing accuracy on large and nonlinear resistive crossbar arrays. 
Users should refer to Fig.\ref{fig_Rw} to see if the linear mapping can provide enough bit-accuracy on the chosen crossbar size and $R_{on}/R_{wire}$ ratio.
If not, the parasitic-aware mapping is encouraged. 
For most cases, especially neural networks, the model parameters only need to map on crossbar once before inference. 
Therefore, using parasitic-aware mapping is a one-time cost and leads to ignorable overhead on inference. 
Row shuffling works on the input stage and also the mapping stage. Output compensation works on the output stage and it tries to compensate for the output error with the inputs. 
Since it involves re-computation of the defects part to improve the VMM accuracy, it becomes less efficient as defect number increases.   

\section{Evaluation}
\subsection{Simulation setup}
\label{sec_simulation}
To measure the performance of our methods, we first check the number of distinguishable levels in analog output and convert it to bit accuracy following eq. \ref{eq_bitacc}:
\begin{equation}
\mathrm{Bit~Accuracy} = log_2(\mathrm{Output~Range}/\mathrm{Avg. Error} + 1)
\label{eq_bitacc}
\end{equation}

\begin{figure*}[tbh]
    \centering
    \includegraphics[ width= \textwidth]{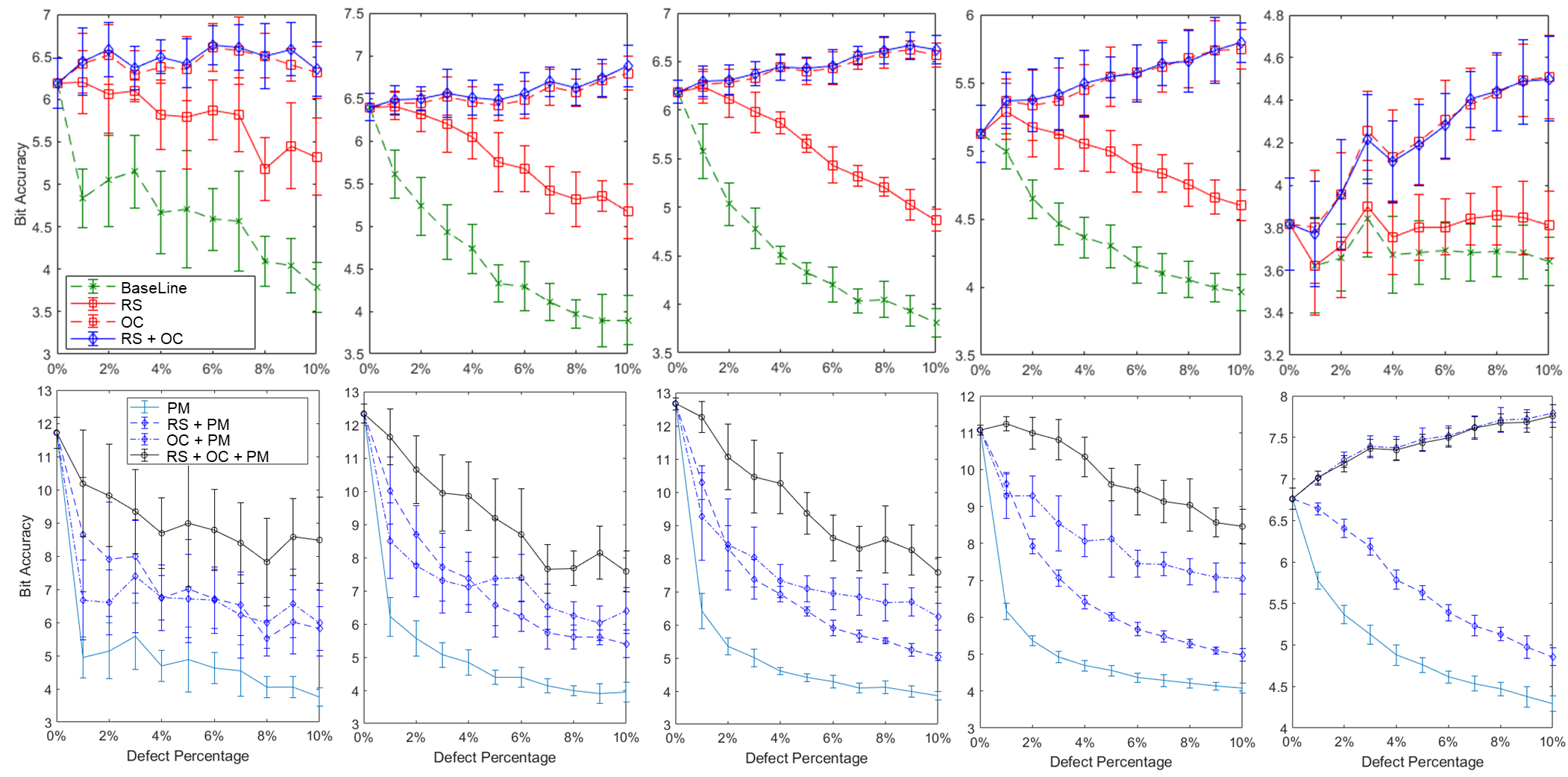}
    \vspace{-20pt} 
    \caption{Bit accuracy of crossbars with different sizes. From left to right, 8$\times$8, 16$\times$16, 32$\times$32, 64$\times$64, 128$\times$128.}
    \vspace{-20pt}
    \label{fig_meanAndworst}
\end{figure*}

In circuit simulation, the memristor model is the TaOx device model \cite{strachan2013state}, and the access transistor model is from \cite{hu2016dot}.
Ron = 15K$\Omega$, Roff = 300K$\Omega$, Rwire = Rin = Rout = 1$\Omega$.
LGS = 1/Roff, and HGS = 1/Ron. 
Our simulation focus on cases that parasitic effects could not be ignored, which is usual in real applications.
For general VMM test, we use uniform-distributed random inputs in range [-1,1], and uniform-distributed random matrices in range [-1,1]. 
For all tests, defect patterns follows a 2-D uniform distribution and then assigned to LGS or HGS with Stuck-ON/Stuck-OFF defect ratio = 1, if not otherwise mentioned.

\begin{figure}[]
    \centering
    \includegraphics[width=0.45\textwidth]{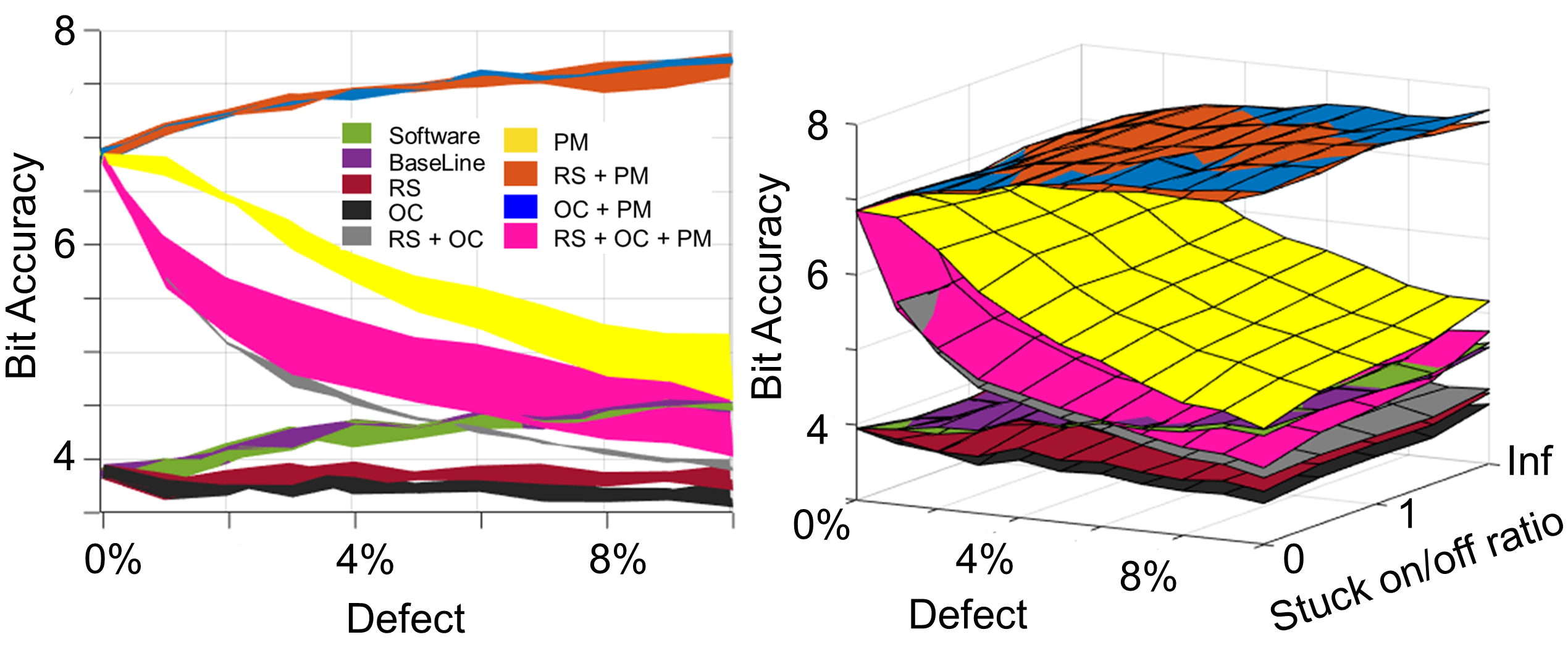}
    \vspace{-9pt} 
    \caption{ Avg. bit accuracy with different Stuck-ON/Stuck-OFF ratios on 128$\times$128 crossbar. \vspace{-2pt}}
    \vspace{-10pt}
    \label{fig_ratio}
\end{figure}

\subsection{General VMM test}

Fig.\ref{fig_meanAndworst} shows the bit accuracy of crossbars with different sizes. Here  
"Baseline" means linear mapping \textbf{G} to defected crossbars, and then do computing without any mitigation.  

The result shows that: 
First, with linear mapping row shuffling provides $\sim$1 bit improvement on average, output compensation provides $\sim$2 bit improvement on high defect rates, and its accuracy increases with defect rate as more computations are transferred to digital circuit implemented compensation component. 
Second, RS+OC performs very similar to output compensation alone in all crossbar sizes and capped at certain bits as defect rate grows. This is because circuit parasitics begin to dominate the error as defects being mitigated. 
OC not only rescues the stuck-at fault but also increases accuracy from the other imperfections. In such parasitics or other defects dominate cases, as the stuck defect goes up, OC could have more input and fitting parameter to compensate output. Therefore, accuracy may also increase with the stuck defect, which usually happens in low bit accuracy, as shown in our results.
Without considering parasitic effects, any mitigation methods for defects will have limited improvement.
Third, with consideration of parasitic-aware mapping(PM) to mitigate parasitic effects, RS+PM,  OC+PM, and RS+OC+PM make a leap in the performance, especially for large crossbars. But only using parasitic-aware mapping without defects mitigation cannot help mitigate the defects impact. 
This supports our claim that parasitic-aware mapping is used to decouple the parasitic impact and boost the performance of row shuffling and output compensation.
RS+OC+PM appears to have the best resilience to high defect rates and can maintain  $\sim$8 bit accuracy in all situations. 
In $128\times 128$ crossbar, OC+PM, and RS+OC+PM capped again since the remaining parasitic effects begin to dominate the error again.
In this case, we need fabrication improvements, such as a device with better linearity and higher resistance range, wires with lower resistance, to overcome the bit accuracy barrier in large crossbars. 

Stuck-ON/Stuck-OFF defect ratio can vary a lot for different chips with different materials, fabrication processes and programming schemes.
Thus, we evaluate the impact of Stuck-ON/Stuck-OFF defect ratio on computing accuracy. 
Fig. \ref{fig_ratio} shows the average bit accuracy with different Stuck ON/Stuck-OFF defect ratio on a 128$\times$ 128 crossbar. We can see that parasitic-aware mapping only mitigates parasitic and suffers a large variation in computing accuracy due to different Stuck-ON/Stuck-OFF defect ratios. Other methods, especially RS+OC+PM, are not sensitive to Stuck-ON/Stuck-OFF defect ratio.

\subsection{CNN demonstration}
\begin{figure}[t]
      \centering
      \includegraphics[width=\columnwidth]{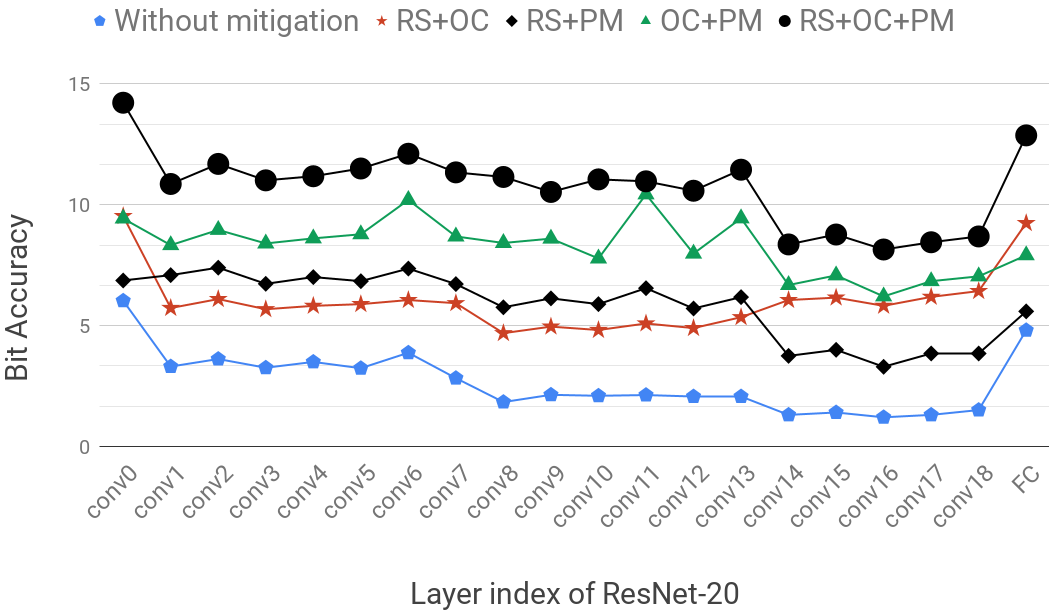}
      \vspace{-20pt} 
      \caption{Bit accuracy of ResNet20 layers.\vspace{-6pt}}
      \vspace{-20pt}
      \label{fig_resnet_bitaccuracy}
\end{figure}

To demonstrate the effeteness of our methods for crossbar-based neural networks, we evaluated them on defected crossbar-based ResNet-20 on CIFAR10 dataset.
We adopted the method in \cite{DBLP:journals/corr/abs-1810-02225} to transform convolution layers to VMM computations.

Fig. \ref{fig_resnet_bitaccuracy} shows the bit accuracy of different combinations in ResNet20 layers with 10\% defects in crossbars. 
It is obvious that RS+OC+PM significantly improves the computng accuracy, especially at high defect rates and large crossbars. 
Bit accuracy between conv14-18 is relatively lower than other layers because the crossbar used in those layers are much larger.
It worth noting that even a defect-free crossbar can only get $\sim$8-bit computing accuracy with PM\cite{hu2016dot}. Thanks to OC, our defected crossbars can get even higher bit accuracy than defect-free crossbars at the cost of digital computing assistance.
Even on the largest crossbar (576$\times$64), our proposed methods could still maintain $>$8-bit output accuracy. 
Our result does not only demonstrate the effeteness of our defect mitigation methods, but also points out a potential direction of crossbar/ALU hybrid computing if more computing accuracy is needed. 

\begin{figure}[t]
      \centering
      \includegraphics[width=0.45\textwidth]{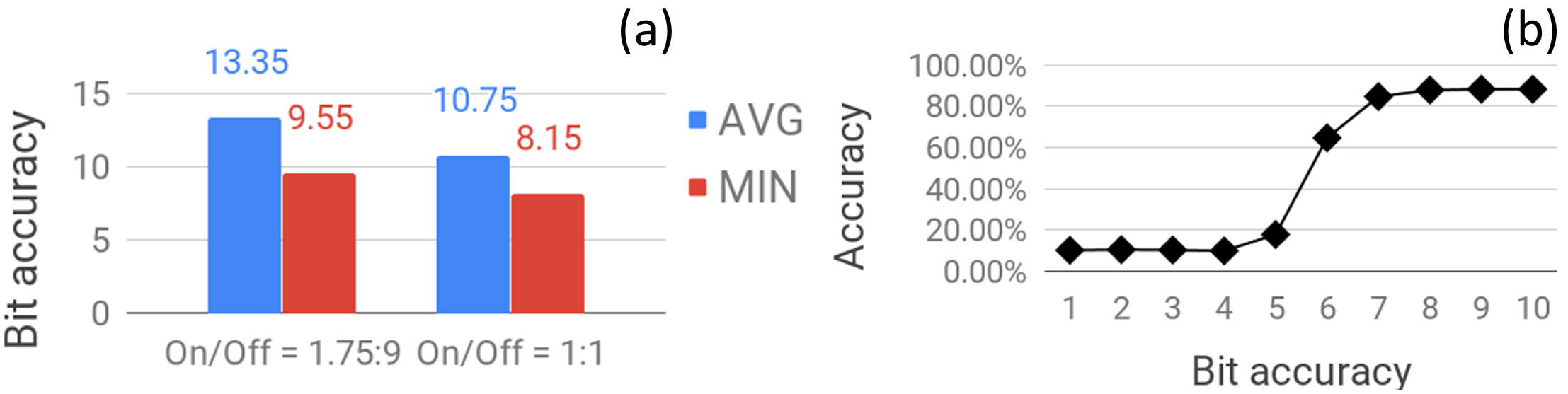}
      \vspace{-9pt} 
      \caption{Bit accuracy of ResNet20 with 10\% defect and different ON/Off ratio.\vspace{-4pt}}
      \vspace{-20pt}
      \label{fig_diff_ratio_resnet}
\end{figure}

Resistive crossbar based CNN is more sensitive to Stuck-ON defect rather than Stuck-OFF\cite{He:2019:NIA:3316781.3317870}. Therefore, lower Stuck-ON/OFF ratio under the same defect rate would cause more error in the CNN application. In Fig. \ref{fig_diff_ratio_resnet}, we tested ResNet-20 with 10\% defects with RS+OC+PM. Two Stuck-ON/Stuck-OFF ratios are tested:  1.75/9 from \cite{6725492} and 1/1.
Fig.\ref{fig_diff_ratio_resnet}(a) shows that higher Stuck-ON/Stuck-OFF ratio leads to lower minimum and average bit-accuracy in all CNN layers. 
Fig.\ref{fig_diff_ratio_resnet}(b) shows that with minimum bit-accuracy $>$ 8-bit, crossbar-based ResNet-20 on CIFAR10 maintains software-level classification accuracy (88.4\%), proves the effeteness of our methods in CNN applications.
\section{Discussion}
\label{sec_discussion}
\subsection{Implementation strategy}
\begin{figure}[t]
      \centering
      \vspace{6pt}
      \includegraphics[width=0.7\columnwidth]{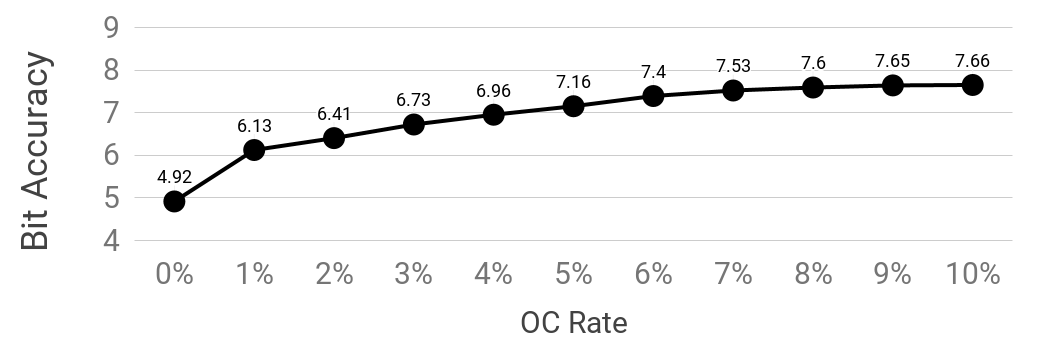}
      \vspace{-9pt}
      \caption{Bit accuracy for limited OC.}
      \vspace{-16pt}
      \label{fig_OC_percent}
\end{figure}
Since the optimal implementation can only be done with architecture level analysis, here we only provide some discussion on implementation strategies, and it is not the focus of the paper.
 
To implement RS in hardware, we need an interconnect  network between data inputs and crossbar rows. 
Every input should be able to connect to any row of crossbar to guarantee arbitrary shuffling order is reachable. 
Such hardware refers to non-blocking interconnect networks \cite{1012681}. 
In our case, as the shuffling order is fixed, dynamic routing configuration is very rare and dynamic power consumption is near zero. 
To implement RS in software, we need re-order input data. 
Re-ordering data in memory with the given order can be quickly done in $O(n)$. 
Due to pipeline processing, software implementation of RS may not affect the throughput but only affects latency since the main bottlenecks are still at DACs and ADCs.
Some memristor-based architectures, such as Atomlayer\cite{Qiao:2018:AUR:3195970.3195998}, store all inputs and intermediate data in DRAM. In this case, software implementation can be performed within the DRAM at low cost. 

To implement OC in hardware, additional MACs are required. 
In our approach, OC is limited to at most 10$\%$ of the total computation. 
This is because if we have to re-calculate more than 10$\%$ of the VMM computation, the benefit of using crossbars will be very marginal. 
Hardware implementation can minimize delay but at the expense of extra area and power consumption.
Fig. \ref{fig_OC_percent} shows the impact of OC rate on computing accuracy of a 128$\times$128 crossbar array with 10\% defects. 1\% OC rate means we only consider 1\% defects with the highest conductance error in each column. 10\% OC rate means we compensate for all  10\% defects. Low OC rate saves computation power but limits the OC performance.

To implement OC in software, the compensation is done by CPU, and it runs in parallel with the crossbars.   

Modern CPUs, such as Intel I3, I5 and I7 series support SSE4 instruction set which contains the specific dot-product instructions(DPPS and DPPD), and could be very efficient in handling OC computations\cite{wiki:sse4}.

The hardware overhead is significant in most cases.
From ISAAC \cite{Shafiee:2016:ICN:3007787.3001139}, a $128 \times 128$ crossbar PE needs 0.157 $mm^{2}$ area and 24 mW power. However, a switch network for $128 \times 128$ needs 18.3 with 0.25um technology \cite{wu20022}.
For the OC, if we add MAC after each column, one MAC may need up to 2.1 $mm^{2}$ and 0.17W under 0.8um technology \cite{553190}. 
As a result, software implementation may be more suitable. Crossbar-based architectures like ISAAC,  PRIME \cite{7551380}, Atomlayer \cite{Qiao:2018:AUR:3195970.3195998} could benefit from our methods when encountering defect issues without any hardware modification.

\subsection{Comparison to related work}

Many works have been published on the defect issue, and their methods can be concluded in two ways. One way is to use redundant hardware where redundant rows and columns in a crossbar are used to replace highly defected ones \cite{liu2018design,kim2018energy,tunali2017permanent}. Another way is to re-train ANNs with consideration of defect patterns \cite{chakraborty2017technology,liu2017rescuing}.

With redundant hardware, crossbar size is usually empirically set to 1.5 times of the matrix size regardless of the hardware cost \cite{tunali2017permanent}.
However, it is still necessary to have a re-routing scheme to route input signal from a defected row to a defect-free row. 
Therefore, the connection/switch network, which transmits the input data to its destination row is also required.
Since the redundant crossbar has 50\% more rows, it may need a larger switch network than our hardware RS implementation. 
In short, our proposed method can be applied with redundant crossbars to further mitigate the impact of defects.

Re-training with consideration of defects sounds promising as it can tolerate high defect rates up to 20\% thanks to the sparse nature of ANNs \cite{liu2017rescuing}. However, it has three limitations: first, they are ANNs specific and not for general VMM; Second, the problem must be trainable, or say, there must be enough training data with labels to re-train the ANN with consideration of defects; Third, circuits with different defect patterns have to be re-trained individually even they have the same weight matrix.
Moreover, there are other overheads for re-training need to be considered in practice, including the extra training circuit, re-training time, and even more time-related defects. Because re-training may update weights more frequently and cause the more time-related defects.

Not only manufacturing defects but also time-related defects can also be rescued with our proposed method or redundancy crossbar. 
After long time use, more defects may appear. We can detect crossbar defects again, and use the new defect pattern to update shuffling order, compensation fitting parameters, and even the conductance matrix. 
Overall, our proposed methods are very flexible and can be used together with existing defect mitigation methods.

\section{Conclusion}
\label{sec_conclusion}
Defects in resistive crossbar significantly degrade its performance in computing.
Previous work using redundant hardware and/or re-training methods have limitations on efficiency and applicable ranges. 
In this paper, we introduce row shuffling and output compensation to mitigate defects in resistive crossbar-based analog VMM computing with consideration of circuit parasitics.
Our methods provide high flexibility for implementation and can be used for general VMM applications.
Extensive circuit simulations have been carried out to verify the performance in different applications.

\end{document}